\documentclass[aps,nofootinbib,twocolumn,floatfix,prd,superscriptaddress]{revtex4} %eqsecnum,
\usepackage{graphicx,amsmath,amssymb,color,bm,hyperref,epsfig,epstopdf}
\def\be{\begin{equation}}
\def\ee{\end{equation}}
\def\bea{\begin{eqnarray}}
\def\eea{\end{eqnarray}}
\newcommand{\vs}{\nonumber\\}
\def\ba#1\ea{\begin{align}#1\end{align}}

\newcommand{\refeq}[1]{Eq.~(\ref{eq:#1})}          
\newcommand{\refeqs}[2]{Eqs.~(\ref{eq:#1})--(\ref{eq:#2})}          
          
\newcommand{\reffig}[1]{Fig.~\ref{fig:#1}}          

\newcommand{\refsec}[1]{\S~\ref{sec:#1}}          
\newcommand{\refapp}[1]{App.~\ref{app:#1}}

\renewcommand{\v}[1]{\mathbf{#1}}
\newcommand{\vx}{\v{x}}
\newcommand{\vk}{\v{k}}

\newcommand{\<}{\langle}
\renewcommand{\>}{\rangle}

\renewcommand{\k}{\kappa}
\renewcommand{\d}{\delta}
\newcommand{\D}{\Delta}

\newcommand{\nhat}{\hat{n}}
\newcommand{\vnhat}{\v{\hat{n}}}

\renewcommand{\l}{\ell}

\newcommand{\zt}{\tilde{z}}
\newcommand{\hp}{h_\parallel}
\newcommand{\chib}{\bar{\chi}}
\newcommand{\chit}{\tilde{\chi}}

\newcommand{\iMpch}{\:h/{\rm Mpc}}
\def\Q{\mathcal{Q}}

\def\M{\mathcal{M}}

\def\W{\mathcal{W}}

\def\nhat{\hat{n}}
\def\vnhat{\hat{\v{n}}}
\def\nhatt{\hat{\tilde{n}}}

\begin{document}

\title{Large-Scale Structure with Gravitational Waves I: Galaxy Clustering}

\author{Donghui Jeong}
\affiliation{Department of Physics and Astronomy, Johns Hopkins University, 3400 N. Charles St., Baltimore, MD 21210, USA}
\author{Fabian Schmidt}
\affiliation{Theoretical Astrophysics,
	California Institute of Technology, 
	Mail Code 350-17, Pasadena, CA  91125, USA}

\begin{abstract}
Observed angular positions and redshifts of large-scale structure tracers 
such as galaxies are affected by
gravitational waves through volume distortion and magnification effects.  
Thus, a gravitational wave background can in principle be probed through 
clustering statistics of large-scale structure.  
We calculate the observed angular clustering of galaxies 
in the presence of a gravitational wave background at linear order 
including all relativistic effects.  
For a scale-invariant spectrum of gravitational waves, the effects
are most significant at the smallest multipoles ($2 \leq \ell \leq 5$),
but typically suppressed by six or more orders of magnitude with
respect to scalar contributions for currently allowed amplitudes of the
inflationary gravitational wave background.  We also discuss the most
relevant second-order terms, corresponding to the distortion of 
tracer correlation functions by gravitational waves.  These provide
a natural application of the approach recently developed in \citet{stdruler}.
\end{abstract}

\date{\today}


\maketitle

%%%%%%%%%%%%%%%%%%%%%%%%%%%%%%%%%%%%%%%%%%%%%%%%%%%%%%%%%%%%%%%%%%%%%%%%%%%
%%%%%%%%%%%%%%%%%%%%%%%%%%%%%%%%%%%%%%%%%%%%%%%%%%%%%%%%%%%%%%%%%%%%%%%%%%%
\section{Introduction}
\label{sec:intro}

The origin of the initial perturbations which gave rise to the structure
in the Universe is one of the most profound questions in cosmology.  
Currently, the most popular scenario is inflation \cite{Starobinsky,Guth}, 
a phase of accelerating expansion in the very early Universe which 
produced seed perturbations as quantum fluctuations 
frozen after exiting the horizon.  
One of the key predictions of inflation is a potentially observable
background of stochastic gravitational waves (GW).  A detection of 
a GW background would allow for a determination of the energy scale of
inflation, and pose a significant challenge to competing scenarios
for the origin of the initial perturbations.  

The polarization of the cosmic microwave background (CMB) is widely
considered to be the most promising probe of the primordial GW background
in the near future
\cite{SelZalPRL,KamionkowskiKosowskyStebbins}.  However, given the profound
impact of a detection, it is worth studying complementary observational
techniques, in order to be able to provide independent confirmation of
a positive result.  
The study of the effect of GW on large-scale structure observables has a 
long history.  
\citet{Linder88} considered the distortion of galaxy correlation functions
and derived an upper limit on the GW background.  \citet{Barkana96} studied
the apparent proper motion of distant objects induced by GW
(see also \cite{BookFlanagan,bharadwaj/sarkar:2009}).  
More recently, \citet{BookEtal} have studied the prospects for 
using gravitational lensing of the CMB for this purpose.  At redshifts
of order $10-200$, the 21cm HI emission from the dark ages has been proposed
as a potentially extremely sensitive probe of a GW background
\cite{MasuiPen,BookMKFS}.  Due to its three-dimensional nature and 
observable structure on much smaller scales, 
the 21cm emission should in principle be able to 
probe GW amplitudes orders of magnitude smaller than the CMB.  
The shear, measured through correlations of galaxy ellipticities,
has also been studied as avenue for detecting a GW background
\cite{DodelsonEtal,Dodelson10}, though these authors have concluded 
that this measurement will likely not be competitive with the CMB
(see also \cite{paperII}).    

The goal of this paper, and its companion \cite{paperII}, is to 
systematically and rigorously derive the GW effects on large-scale structure 
observables.  While we restrict ourselves to a linear treatment in the 
tensor perturbations, we strive to keep the results as general as possible
otherwise. 
This paper deals with observed densities of large-scale 
structure tracers, which have so far not been 
investigated in the context of two-point statistics as a probe of GW. 
The companion paper deals with shear (as measured from, e.g., galaxy ellipticities).  

Since there is no 3-scalar that can be
constructed from tensor perturbations at linear order without making
reference to some external (3-)vector or tensor, the ``intrinsic''
density of tracers, i.e. the density that would be measured by a local
comoving observer, is not affected by tensor modes at linear order.  
Thus, the impact of tensor perturbations is exclusively due to projection effects 
which can be derived 
in analogy to \cite{yoo/etal:2009,challinor/lewis:2011,bonvin/durrer:2011,gaugePk} using 
the geodesic equation.  The main observable we consider is the
angular power spectrum $C(l)$ of tracers.  Since the GW effects are most
important on the very largest scales, the 3D power spectrum $P(k)$ is
not a meaningful quantity for this purpose.

Note that in contrast, the shear is itself
a tensorial quantity, and thus there is a possible ``intrinsic'' contribution 
correlated with the GW background, analogous to the intrinsic alignment 
effect present
for scalar perturbations.  This issue, which has not been investigated
before, is the topic of the companion paper \cite{paperII}.

As emphasized by \citet{KaiserJaffe},
there are some key differences in how tensor perturbations
affect photon geodesics
as opposed to scalar perturbations:  scalar modes which are transverse to the
line of sight lead to a significantly amplified coherent deflection, whereas 
the same does not
happen for tensor modes as they themselves propagate at the speed
of light.  Furthermore, while scalar modes grow (in the matter-dominated
era), tensor modes redshift away.  Thus, the intuition that transverse
modes at lower redshifts contribute most of the projection (lensing) effects does
not hold anymore for tensor modes.  Rather, the GW contribution to
the clustering of LSS tracers is
dominated by contributions close to the time of emission, and
time-derivatives of the perturbations are (at least) as relevant
as spatial derivatives.  

The outline of the paper is as follows.  We introduce our notation
and conventions in \refsec{prelim}.  \refsec{proj} presents the
geodesic equation for a tensor mode and the derivation of the
tensor mode contributions to the observed galaxy density, 
including the magnification bias
contribution.  The galaxy angular power spectrum is discussed in 
\refsec{Cl}.  We also highlight the differences to the scalar case
in this calculation.  \refsec{2nd} deals with the relevant higher order
terms neglected in \refsec{proj}.  We conclude in \refsec{concl}.  
The appendix contains details on some aspects of the calculation.

%%%%%%%%%%%%%%%%%%%%%%%%%%%%%%%%%%%%%%%%%%%%%%%%%%%%%%%%%%%%%%%%%%%%%%%%%%%
%%%%%%%%%%%%%%%%%%%%%%%%%%%%%%%%%%%%%%%%%%%%%%%%%%%%%%%%%%%%%%%%%%%%%%%%%%%
\section{Preliminaries}
\label{sec:prelim}

We begin by introducing our convention for metric and tensor perturbations
and some notation.  
For simplicity, we restrict 
ourselves to a spatially flat FRW background, and consider only tensor
(gravitational wave) modes in the main part of the paper.  The perturbed 
metric is then given by
\be
ds^2 
=
a^2(\eta)
\left[
-d\eta^2 + \left(\delta_{ij}+h_{ij} \right) dx^idx^j
\right],
\label{eq:metric}
\ee
where $h_{ij}$ is a metric perturbation which is 
transverse and traceless:
\be
h^i_{\; i} = 0 = (h_{ik})^{,i}.
\ee
In order to simplify the analysis, we shall also consider the conformal metric,
\be
d\bar{s}^2 
=
-d\eta^2 + \left(\delta_{ij}+h_{ij} \right) dx^idx^j,
\label{eq:metric_conf}
\ee
where $\eta$ denotes the conformal time, for our analysis of the light deflection.

We then decompose $h_{ij}$ into Fourier modes of two polarization states,
\ba
h_{ij}(\vk, \eta) =\:& e^+_{ij}(\hat\vk) h^+(\vk,\eta) + e^\times_{ij}(\hat\vk)
h^\times(\vk, \eta),
\label{eq:hpol}
\ea
where $e^s_{ij}(\hat\vk)$, $s=+,\times$, are transverse 
(with respect to $\hat\vk$) and 
traceless polarization tensors normalized through
$e^s_{ij} e^{s'\:ij} = 2 \d^{ss'}$.  We assume both polarizations to
be independent and to have equal power spectra:
\ba
\< h_{s}(\vk,\eta) h_{s'}(\vk',\eta') \> =\:& (2\pi)^3 \d_D(\vk-\vk') 
\d_{ss'} \frac14 P_T(k,\eta,\eta').
\label{eq:PT}
\ea
Here, $\eta$ denotes conformal time, and the unequal-time power spectrum
is given by
\ba
P_{T}(k,\eta,\eta') =\:& T_T(k,\eta) T_T(k,\eta') P_{T0}(k),
\label{eq:PT2}
\ea
where $T_T(k,\eta)$ is the tensor transfer function, and the primordial
tensor power spectrum is specified through an amplitude $\Delta_T^2$ and
an index $n_T$ via
\ba
P_{T0}(k) =\:& 2\pi^2\,k^{-3} \left(\frac{k}{k_0}\right)^{n_T} \Delta_T^2.
\label{eq:PT0}
\ea
Following \emph{WMAP} convention \cite{komatsu/etal:2011},
we choose $k_0 = 0.002\:{\rm Mpc}^{-1}$ as pivot scale.  Throughout,
we will assume a scalar-to-tensor ratio of $r = 0.2$ at $k_0$ (consistent with
the 95\% confidence level WMAP bound), which together with our fiducial
cosmology determines $\Delta_T^2$.  The tensor index is chosen to follow
the inflationary consistency relation, $n_T = -r/8 = -0.0025$.  For the
expansion history,  we assume
a flat $\Lambda$CDM cosmology with $h=0.72$ and $\Omega_m=0.28$.  Contributions
from scalar perturbations are evaluated using a
spectral index of $n_s=0.958$ and power spectrum normalization at $z=0$ of
$\sigma_8 = 0.8$.

From \refeq{hpol} and \refeq{PT}, we easily obtain
\ba
\< h_{ij}(\vk,\eta) h_{kl}(\vk',\eta') \> =\:& (2\pi)^3 \d_D(\vk-\vk') \\
& \times\left[ e^+_{ij}(\hat\vk) e^+_{kl}(\hat\vk) + e^\times_{ij}(\hat\vk) e^\times_{kl}(\hat\vk) \right] \vs
& \times \frac14 P_T(k,\eta,\eta') \vs
\< h_{ij}(\vk,\eta) h^{ij}(\vk',\eta') \> =\:& (2\pi)^3 \d_D(\vk-\vk') P_T(k,\eta,\eta'). \nonumber
\ea
Long after recombination, the transverse anisotropic stress which sources 
gravitational waves becomes negligible, and the tensor modes propagate
as free waves.  During matter-domination, the tensor 
transfer function then simply becomes
\be
T_T(k,\eta) = 3 \frac{j_1(k \eta)}{k\eta},
\label{eq:TT}
\ee
which however is still valid to a high degree of accuracy during the current 
epoch of acceleration.  We will use \refeq{TT} throughout.

%%%%%%%%%%%%%%%%%%%%%%%%%%%%%%%%%%%%%%%%%%%%%%%%%%%%%%%%%%%%%%%%%%%%%%%%%%%
%%%%%%%%%%%%%%%%%%%%%%%%%%%%%%%%%%%%%%%%%%%%%%%%%%%%%%%%%%%%%%%%%%%%%%%%%%%
\section{Tensor contributions to the observed galaxy density}
\label{sec:proj}

In this section we derive the GW contribution to the 
observed density of tracers, including the magnification bias effect.  
We follow the notation in \citet{gaugePk} (see also \cite{stdruler}) which is 
summarized in their Sec.~II~A.  
The zero-th order photon geodesic in conformal coordinates 
[\refeq{metric_conf}] 
is simply
\be
\bar{x}^\mu(\chi)
=
\left(
\eta_0-\chi, \vnhat\chi
\right),
\label{eq:geod_conf}
\ee
where the comoving distance $\chi$ along the geodesic serves as
affine parameter, with $\chi=0$ corresponding to the observer's location.  
Here and throughout, $\vnhat$ denotes the unit
vector in the direction of the \emph{observed} position of the source
($\nhatt$ in \cite{gaugePk}).  Hence,
\be
\frac{d\bar x^\mu}{d\chi} = (-1, \vnhat).
\ee
In the following, $\zt$ will stand for the observed redshift of the
source, and $\chit \equiv \bar\chi(\zt)$ is the conformal distance
corresponding to that redshift when evaluating the distance-redshift
relation $\bar\chi(z)$ in the background.  

We decompose vectors into transverse and longitudinal parts with
respect to the line-of-sight,
\ba
X_\parallel \equiv\:& \nhat_i\,X^i\vs
X_\perp \equiv\:& X^i - \nhat^i\, \nhat_j\, X^j,
\ea
and correspondingly define longitudinal and transverse derivatives through
\ba
\partial_\parallel \equiv\:& \nhat^i\,\partial_i \vs
\partial_\perp^i \equiv\:& \partial^i - \nhat^i \partial_\parallel.
\ea
Finally, we define
\be
\nabla^2_\perp \equiv
\partial_{\perp i}\partial_\perp^i
=\nabla^2 - \partial_\parallel^2 - \frac{2}{\chit}\partial_\parallel,
\ee
and make use of 
\ba
\partial_\parallel \nhatt^i =\:& \nhatt^i\partial_{\perp i} = 0.
\ea

%%%%%%%%%%%%%%%%%%%%%%%%%%%%%%%%%%%%%%%%%%%%%%%%%%%%%%%%%%%%%%%%%%%%%%%%%%
\subsection{Photon geodesics with a tensor mode}

We now briefly outline the derivation of the displacements $\D x^i$ of the
true emission point from the observationally inferred position.  This
is a special case of the derivation in App.~B of \citet{stdruler}, to which
the reader is referred for more details.  
We parameterize the linear order deviation of the 
photon geodesic as
\be
\frac{dx^\mu}{d\chi} = (-1 + \delta\nu,\vnhat+\delta\v{e}).
\ee
The initial conditions for integrating the geodesic
equation are set by demanding that the components of the photon momentum
measured in a locally orthonormal frame at the observer's location match
$\vnhat$.  
For that, we construct an orthonormal tetrad $(e^a)_\mu$ ($a=0,1,2,3$) 
carried by an observer so that at the observer's location
\be
g^{\mu\nu}(e^a)_\mu(e^b)_\nu = \eta^{ab}.
\ee
Then, the photon four-momentum measured by the observer is given by
\be
\left(1,\nhat^i\right) = 
\left((e^0)_\mu, (e^i)_\mu\right) \frac{dx^\mu}{d\chi}.
\ee
A detailed calculation is presented in appendix B of \cite{stdruler}.  
Using that the four-velocity of comoving observers is given by
$u^\mu = a^{-1}(1,0,0,0)$, this leads to
\ba
\d\nu(\chi=0) =\:& 0\vs
\d e^i(\chi=0) =\:& -\frac12 (h^i_{\;j})_o\: \nhat^j.
\ea
Here and throughout, a subscript $o$ indicates that the quantity is 
evaluated at the observer's location.  
The corresponding initial condition for the geodesic equation for comoving observers
with general metric is presented in Eq. (B11) of \cite{stdruler}.  
In case of scalar perturbations, including $\d e^i(\chi=0)$ is important to
ensure gauge-invariant expressions \cite{gaugePk}.  While there
is no gauge ambiguity in tensor modes, we will show that the observer
term is numerically important for the quadrupole of the observed
galaxy density.  
The redshift perturbation is related to the shift in the frequency through
\be
\delta z(\chit)
= -\delta \nu = \frac{1}{2} \int_0^{\chit}h_\parallel'd\chi,
\ee
where we have defined
\be
h_\parallel \equiv h_{ij}\,\nhat^i\nhat^j.
\ee
Here and hereafter, a prime denotes a derivative with respect to conformal time,
if not used for distinguishing different variables. The distinction between
the two should be clear from the context.  
The redshift of the photon along the perturbed geodesic is given by
$1+z(\chi) = [1+\d z(\chi)]/a(x^0(\chi))$.  Requiring that the 
redshift at emission equals $\zt$ yields an equation for the
first-order perturbation to the affine parameter at emission $\chi_e$ 
\cite{gaugePk},
\ba
\chi_e =\:& \chit + \d\chi \vs
\d\chi =\:& \d x^0 - \frac{1+\zt}{H(\zt)}\d z,
\ea
where $\d x^0$ is the perturbation to the 0th component of the geodesic
evaluated at $\chi=\chit$.  
We now relate the observed position $\tilde{\vx}$, inferred assuming 
the unperturbed geodesic $\bar{x}^\mu$, and the true position $\vx$ through
(see Fig.~1 in \cite{gaugePk})
\be
\Delta \vx \equiv 
\vx - \tilde{\vx} = \d x(\chit) + \vnhat\d \chi,
\ee
where $\d x(\chit)$ is the spatial perturbation to the geodesic evaluated
at $\chi=\chit$.  
We can then decompose the displacement $\D \vx$ into perpendicular
and longitudinal parts,
\ba
\Delta x_\parallel
=\:& \delta x^i \nhat_i + \delta x^0 - \frac{1+\zt}{H(\zt)}\delta z
\\
\Delta x_\perp =\:& \Delta x^i - \nhat^i\D x_\parallel.
\ea
Explicitly, 
\ba
\Delta x_\parallel =\:& - \frac{1}{2}
\int_0^{\chit} d\chi \: h_\parallel
-\frac{1+\zt}{2H(\zt)}
\int_0^{\chit}d\chi\: h_\parallel'
\label{eq:Dxpar}
\ea
and 
\ba
\Delta x_\perp^i
=\:&
\frac{1}{2}\chit
\left[
\left(h_{ij}\right)_o\nhat^j - \left(h_\parallel\right)_o \nhat^i
\right]
\vs
&+
\int_0^{\chit}
d\chi
\left\{
\frac{\chit-\chi}{2}
\partial_\perp^i
h_{\parallel}
+
\frac{\chit}{\chi}
\left(
h_\parallel \nhat^i - h_{ij}\nhat^j
\right)
\right\}.
\label{eq:Dxperp}
\ea
Note that \refeqs{Dxpar}{Dxperp} can be obtained directly from
Eqs.~(43)--(47) in \cite{stdruler} (with $\d z = \D\ln a$) by setting
$A = B = v = 0$, or, alternatively, from
the corresponding Eqs.~(38)--(39) in \cite{gaugePk} by restricting
to a transverse-traceless metric perturbation and setting
$E_\parallel = h_\parallel/2$.   

In the following, we will further need the \emph{convergence} $\hat\k$,
defined through (see \refapp{kappa})
\ba
\hat\k \equiv\:& -\frac{1}{2}\partial_{\perp i}\Delta x_\perp^i \vs
=\:& \frac54 h_{\parallel o} -\frac12 h_{\parallel}
 - \frac12 \int_0^{\chit} d\chi \Big[ 
h_\parallel' + \frac3{\chi} h_\parallel \Big] \vs
 & - \frac14 \nabla^2_\Omega \int_0^{\chit} d\chi \frac{\chit-\chi}{\chit\:\chi} 
h_\parallel. \label{eq:kappa}
\ea
Here $\nabla^2_\Omega = \chi^2 \nabla_\perp^2$ is the Laplacian on the
sphere.  
Also, we will use
\ba
\partial_{\chit} \Delta x_\parallel =\:&
-\frac{1}{2} h_\parallel
-
\frac{1+\zt}{2H(\zt)} h_\parallel' \vs
& - \frac{H(\zt)}{2} 
\frac{\partial}{\partial \zt}
\left[
\frac{1+\zt}{H(\zt)}
\right] 
\int_0^{\chit} d\chi h_\parallel',
\ea
where we have used $d\chit = d\zt/H(\zt)$.
While $\hat\kappa$ is the usual coordinate convergence,
$\partial_{\chit} \Delta x_\parallel $ is the distortion of the volume
along the line of sight and can thus be seen as a ``radial convergence''.

%%%%%%%%%%%%%%%%%%%%%%%%%%%%%%%%%%%%%%%%%%%%%%%%%%%%%%%%%%%%%%%%%%%%%%%%%%
\subsection{Observed galaxy density}

The observed comoving number density of galaxies $a^3 \tilde n_g$ is related to 
the \textit{true} comoving number density $a^3 n_g$ through
\be
a^3(\zt) \tilde{n}_g(\tilde{\bm{x}},\zt)
=
\left(
1+\frac{1}{2}\d g^\mu_{\; \mu}
\right)
a^3(\bar{z}) n_g(\bm{x},\bar{z})
\left(
1+\frac{\partial \Delta x^i}{\partial \tilde{x}^i}
\right).
\label{eq:cov_dens}
\ee
The first term in brackets comes from the covariant volume factor $\sqrt{|g|}$
and is equal to unity, since for transverse-traceless metric perturbations 
$\d g^\mu_{\;\mu}=0$ at linear order.  The
factor $a^3(\bar z) n_g(\vx,\bar z)$ is the true comoving number density
at the point of emission, which we expand as
\be
a^3(\bar z) n_g(\vx, \bar z) = a^3(\bar z) \bar n_g(\bar z)\:[1 + \d_g(\vx, \bar z)],
\label{eq:ngcom}
\ee  
by defining the intrinsic perturbations to the \emph{comoving} number density
$\d_g$.  $\bar z$ is the redshift that would be measured
for the source in an unperturbed universe, and is related to $\zt$ through
\be
1 + \zt = (1+\bar z)(1+\d z).
\ee
Note that, when inserting \refeq{ngcom} into \refeq{cov_dens}, the distinction between $\vx$ and $\tilde\vx$ in the argument
of $\d_g$ is second order if we regard intrinsic galaxy density perturbations
as first order, which we will do in this section.  The relevant additional
terms will be studied in \refsec{2nd}.  
Finally, $1 + \partial_i \D x^i$ is the volume distortion due to gravitational
waves, which as derived in \cite{gaugePk} becomes
\be
\frac{\partial \Delta x^i}{\partial \tilde{x}^i}
=
\partial_{\chit} \Delta x_\parallel
+
\frac{2\Delta x_\parallel}{\chit} - 2 \hat\kappa.
\label{eq:jacobian}
\ee
Thus, gravitational waves affect the observed density of galaxies through
a volume distortion effect, and by
perturbing their redshifts so that we compare the measured galaxy
density $\tilde n_g$ to the ``wrong'' background density $\bar n_g(\zt)$.  
The latter effect is quantified by the parameter
\be
b_e \equiv \frac{d\ln (a^3\bar n_g)}{d\ln a}\Big|_{\!\zt} 
= - (1+\zt)\frac{d\ln (a^3\bar n_g)}{dz}\Big|_{\!\zt}.
\label{eq:btdef}
\ee
Note that this parameter can be measured for a given galaxy sample,
provided the redshift-dependence of the selection function is understood.  

We can now summarize the tensor contributions to the observed galaxy
density perturbation as
\ba
\tilde\delta_{gT}(\tilde{\bm{x}},\zt)
=\:& b_e\delta z - 2\hat\kappa
- \frac{1}{\chit}
\left[
\int_0^{\chit} d\chi h_\parallel
+
\frac{1+\zt}{H(\zt)}
\int_0^{\chit} d\chi h_\parallel'
\right]
\vs
- & \frac{1}{2} h_\parallel
-
\frac{1+\zt}{2H(\zt)} h_\parallel'
-
\frac{H(\zt)}{2} 
\frac{\partial}{\partial \zt}
\left[
\frac{1+\zt}{H(\zt)}
\right] 
\int_0^{\chit} d\chi h_\parallel'.
\label{eq:dT1}
\ea
Here, the subscript $T$ denotes tensor contributions, and we have assumed 
that the intrinsic density perturbation $\d_g$
does not correlate with the tensor modes (following the arguments
in \refsec{intro}).  
%We will return to this issue in \refsec{concl}.  
Note that $h_{ij}$ only enters through $h_\parallel$.  This has to be the
case, since the galaxy density is a scalar quantity and $h_\parallel$ is the 
only non-trivial scalar linear in $h_{ij}$.  The latter also implies that
tensor modes do not contribute to the monopole and dipole of the galaxy
density.  

% % % % % % % % % % % % % % % % % % % % % % % % % % % % % % % % % % % %
\subsection{Magnification bias}
\label{sec:magn}

In the last section, we have assumed that all galaxies are included
in the sample.  In reality, most large-volume surveys are limited in
flux.  A cut on observed flux induces additional fluctuations in the 
galaxy density, since perturbations to the photon geodesic (e.g.,
gravitational lensing) modify the observed flux of a given source.  
The magnification $\M$ can be derived as the perturbation to the
angular diameter distance squared (\cite{gaugePk}, with $\d\M_{\rm there} = \M_{\rm here}$),
\be
1+\M \equiv \frac{\bar{D}^2_A(\zt)}{D^2_A},
\ee
where $D_A$ is the true angular diameter distance to the source while
$\bar D_A(z)$ is the background angular diameter distance-redshift relation.  
Alternatively, one can use the standard ruler approach of
\cite{stdruler}, which for a metric of the form \refeq{metric}
yields (see Eq.~(105) in \cite{stdruler}, where we set $d\ln r_0/d\ln a =0$
neglecting the higher order effect of an evolving number count slope)
\be
\M_T =
- 2\d z +\frac{1}{2}h_\parallel
- \frac{2\Delta x_\parallel}{\chit} + 2 \hat\kappa.
\label{eq:dM}
\ee
We then parametrize the effect on the observed galaxy density through
a parameter $\Q$,
\be
\tilde\delta_{gT}  = \tilde\delta_{gT}(\mathrm{no~magn.}) + \Q \M_T.
\ee
For a purely flux-limited survey, $\Q = - d\ln \bar{n}_g/d\ln f_{\rm cut}$,
where $f_{\rm cut}$ is the flux cut.  More generally, $\Q$ can also
receive a contribution from a size cut \cite{schmidtetal09}.  

% % % % % % % % % % % % % % % % % % % % % % % % % % % % % % % % % % % %
\subsection{Summary of tensor contributions}
\label{sec:dg}

Combining the result of the last two sections, we obtain the
following expression for the linear-order tensor contributions
to the observed galaxy density:
\ba
\tilde\delta_{gT}
=\:&
(b_e - 2\Q )\delta z
-2(1-\Q)\hat\kappa
-\frac{1-\Q}{2} h_\parallel
-
\frac{1+\zt}{2H(\zt)} h_\parallel'
\vs
&-
\frac{1-\Q}{\chit}
\left[
\int_0^{\chit} d\chi h_\parallel
+
\frac{1+\zt}{H(\zt)}
\int_0^{\chit} d\chi h_\parallel'
\right]
\vs
&
-
\frac{H(\zt)}{2} 
\frac{\partial}{\partial \zt}
\left[
\frac{1+\zt}{H(\zt)}
\right] 
\int_0^{\chit} d\chi h_\parallel'.
\label{eq:dgT}
\ea
For later convenience, we reorder the terms as follows:
\ba
\tilde\d_{gT} 
=\:& f_{\chit} \hp + f_{\chit}' \hp' + f \int \frac{d\chi}{\chi} \hp
+ \tilde f \int \frac{d\chi}{\chit} \hp \label{eq:dgT2} \\
& + f' \int d\chi \: \hp'
+ f_\k \nabla^2_\Omega \int d\chi\frac{\chit-\chi}{\chi\,\chit} \hp
+ f_o\, h_{\parallel o}.\nonumber
\ea
Here, all terms outside integrals (without subscript $o$) 
are evaluated at $\chit$, and the
integrals go from 0 to $\chit$.  The coefficients are given by
\ba
f_{\chit} =\:& -\frac12 (\Q - 1) \label{eq:coeff}\\
f'_{\chit} =\:& - \frac{1+\zt}{2H} \vs
f =\:& -3 (\Q-1) \vs
\tilde f =\:&  \Q-1 \vs
f' =\:&  \frac12 \left(b_e -1 -2\Q + (1+\zt)\frac{dH/d\zt}{H}\right)
- (\Q-1) \vs
& + (\Q-1) \frac{1+\zt}{H\chit}
 \vs
f_\kappa =\:& -\frac12 (\Q-1)\vs
f_o =\:& \frac52 (\Q-1)
.\nonumber
\ea

%%%%%%%%%%%%%%%%%%%%%%%%%%%%%%%%%%%%%%%%%%%%%%%%%%%%%%%%%%%%%%%%%%%%%
%%%%%%%%%%%%%%%%%%%%%%%%%%%%%%%%%%%%%%%%%%%%%%%%%%%%%%%%%%%%%%%%%%%%%
\section{Observed galaxy power spectrum}
\label{sec:Cl}

%%%%%%%%%%%%%%%%%%%%%%%%%%%%%%%%%%%%%%%%%%%%%%%%%%%%%%%%%%%%%%%%%%%%%
\subsection{Angular power spectrum}
\label{sec:ClA}

\begin{figure}[t!]
\centering
\includegraphics[width=0.5\textwidth]{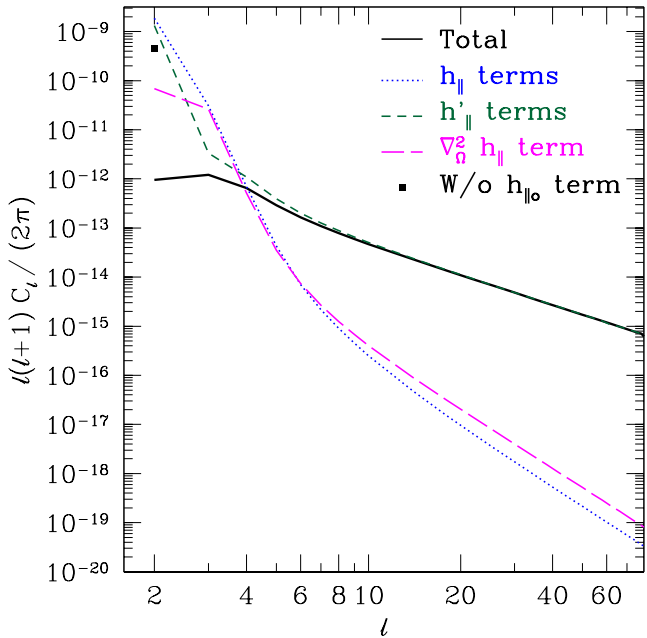}
\caption{Contributions to the observed galaxy angular power spectrum
from inflationary gravitational waves, for a sharp source galaxy redshift
of $\zt = 2$, and using the tensor mode power spectrum defined in
\refsec{prelim}.    
The black solid line shows the total contribution, while
the colored lines show contributions proportional to line-of-sight integrals
of $h_\parallel$ (blue dotted), $h_\parallel'$ (green short-dashed) and 
$\nabla_\Omega^2 h_{\parallel}$ (magenta long-dashed).  Here, we have
assumed $b_e = 2.5,\; Q = 1.5$.  The black square at $l=2$ indicates
the result for $l=2$ if the observer term is neglected (see \refsec{quad}).   
}
\label{fig:Cl_TT}
\end{figure}

Consider a galaxy sample with a redshift distribution $dN/dz$, normalized
to unity in redshift.  Then, the projected galaxy overdensity as a function of
position on the sky is given by
\be
\D_g(\vnhat) = \int_0^\infty d\zt \frac{dN}{d\zt} \tilde \d_g(\bar\chi(\zt) \vnhat; \zt),
\ee
We will assume that the quantities $b_e,\:\Q$ describing the galaxy sample 
are independent of redshift for simplicity.  
We can then write the multipole coefficients of the galaxy density as
\ba
a^g_{lm} =\:& \int d^2\vnhat\: Y^*_{lm}(\vnhat) \D_g(\vnhat).
\ea
We can write all individual contributions to \refeq{dgT} as
\ba
A(\vnhat, \chit) =\:& \int_0^{\chit} d\chi \: W_A(\chi,\chit) h_\parallel(\chi\vnhat,\chi) \vs
=\:& \int_0^{\chit} d\chi \: W_A(\chi,\chit) \vs
&\times \int \frac{d^3\vk}{(2\pi)^3}
e^{i\vk\cdot\vnhat \chi} \nhat^i \nhat^j h_{ij}(\vk,\eta_0-\chi).
\ea
Note that terms 
involving $h_\parallel'$ can be brought into the form $A(\vnhat)$ by
including $d\ln T_T(k,\eta)/d\eta$ in $W_A(\chi)$.  The observer
term $5h_{\parallel o}/3$ contained in $\hat\k$ can similarly be written
with $W_A(\chi) = 5/3\,\d_D(\chi)$.  We will deal with that term specifically
in \refsec{quad}.  By changing the order
of integration, we can then write the contribution
to the projected galaxy overdensity as
\ba
A(\vnhat) =\:& \int_0^\infty d\zt \frac{dN}{d\zt} A(\vnhat,\chit) 
 = \int_0^\infty d\chi\: \W_A(\chi) h_\parallel(\chi\vnhat, \chi) \vs
\W_A(\chi) \equiv\:& \int_{z(\chi)}^\infty d\zt \frac{dN}{d\zt} W_A\left(\chi,\bar\chi(\zt)\right).
\label{eq:WW}
\ea
Note that if $W_A = \d_D(\chi-\chit)$, $\W_A(\chi) = (H dN/dz)|_{z(\chi)}$.  
We now consider the contribution of a single plane wave tensor
perturbation with $\vk$-vector aligned with the $z$-direction.  
Then,
\ba
\nhat^i \nhat^j h_{ij}(\vk,\eta) =\:& \sin^2\theta 
\left[\cos 2\phi\:  h^+(\vk,\eta) + \sin 2\phi\: h^\times(\vk,\eta)\right] \vs
=\:& \sin^2 \theta \left[e^{i2\phi} h_1  + e^{-i2\phi} h_2 \right],
\ea
where
\ba
h_{1,2} \equiv\:& \frac12 (h^+ \pm i h^\times).
\ea
Note that the power spectra of these circular polarization staters are
$P_{h_1h_2} = P_{h_2h_2} = P_T/8$, while $P_{h_1h_2} = 0$.  
Let us denote as $A(\vnhat, \vk)$ the contribution to $A(\vnhat)$ 
from this plane-wave tensor perturbation.  We have
\ba
A(\vnhat, \vk) =\:& \int d\chi\:\W_A(\chi) e^{i k\chi \mu} 
(1-\mu^2) \\
& \times\left[e^{2i\phi} h_1(\vk,\eta_0-\chi) + e^{-2i\phi} h_2(\vk,\eta_0-\chi)\right] \nonumber
\ea
where $\mu = \cos\theta$ is the cosine of the angle between $\vnhat$
and $\hat\vk$.  Note the $e^{\pm 2i\phi}$ factors which are the
key difference to the case of scalar perturbations.  The multipole
coefficients of $A$ are then obtained as follows:
\be
a^A_{lm} = \int \frac{d^3\vk}{(2\pi)^3} a^A_{lm}(\vk),
\ee
\begin{widetext}
\ba
a^A_{lm}(\vk) =\:& \int d^2\vnhat\: Y^*_{lm}(\vnhat) A(\vnhat) \vs
=\:& \int d\chi\:\W_A(\chi) \int d^2\vnhat\: Y^*_{lm}(\mu,\phi) e^{i k\chi \mu} 
(1-\mu^2) \left[e^{2i\phi} h_1(\vk,\eta_0-\chi) + e^{-2i\phi} h_2(\vk,\eta_0-\chi)\right].
\ea
We now use the relation (see App.~A1 in \cite{stdruler})
\ba
\int d\Omega\: Y^*_{lm} (1-\mu^2) e^{\pm i 2 \phi} e^{i x \mu} 
=\:& -\sqrt{4\pi (2l+1)} \sqrt{\frac{(l+2)!}{(l-2)!}}\, i^l \frac{j_l(x)}{x^2} \d_{m \pm 2},
\ea
which yields
\ba
a^A_{lm}(\vk) 
=\:& -\sqrt{\frac{2l+1}{4\pi}\frac{(l+2)!}{(l-2)!}} (4\pi) i^l
\int d\chi\:\W_A(\chi) \left[h_1(\vk,\eta_0-\chi)\d_{m2} + h_2(\vk,\eta_0-\chi)\d_{m-2}\right]
\frac{j_l(k\chi)}{(k\chi)^2}.
\ea
\end{widetext}
Further, we can use the properties of the
spherical harmonics to obtain
\be
a_{lm}[\nabla^2_\Omega A] = - l(l+1) a^A_{lm},
\ee
which we will use to evaluate the convergence contribution.  
Using the definition of the angular power spectrum, we can now
easily write down the cross-correlation between two different 
projections $A,\,B$ of $h_\parallel$ (App.~A1 in \cite{stdruler}):
\ba
C^{AB}_l \equiv\:& \frac{1}{2l+1} \sum_m \mbox{Re} \< a^{A*}_{lm} a^B_{lm} \> \vs
=\:& \frac1{2\pi} \frac{(l+2)!}{(l-2)!}
\int k^2 dk\: P_{T0}(k) F_l^A(k) F_l^B(k) \label{eq:CABl} \\
F_l^X(k) \equiv\:& \int d\chi\:\W_X(\chi) T_T(k,\eta_0-\chi) \frac{j_l(k\chi)}{(k\chi)^2}.\label{eq:Fl}
\ea
It is then straightforward to evaluate auto-
and cross-correlations of the observed 
angular correlation of galaxies, neglecting the contribution of the
term $h_{\parallel o}$ for the moment (see \refsec{quad}).    
In particular, if we want to
evaluate the total tensor contribution to the angular
(auto-)power spectrum of galaxies, we set $W_A = W_B = W_g$, where
\ba
W_g(\chi) =\:& f_{\chit} \d_D(\chi-\chit) + 
\frac{d\ln T_T}{d\eta}\Big|_{\eta_0-\chi} \left( f'_{\chit}  \d_D(\chi-\chit)
+ f'\right) \vs
& + f \frac1\chi + \tilde f \frac1{\chit} - f_\k \:l(l+1) \frac{\chit-\chi}{\chi\,\chit},
\label{eq:Wg}
\ea
and the coefficients are defined in \refeq{coeff}.  Note that the 
divergent pieces $\propto \chi^{-1}$ cancel for $l=2$:
\be
f - 6 f_\k = 0,
\ee
while for $l \geq 3$ the Bessel function in \refeq{Fl} ensures that all terms
are regular.  \reffig{Cl_TT} shows
numerical results for a galaxy sample with a sharp (observed) redshift of $\zt=2$, 
and for $b_e = 2.5,\; Q=1.5$.  The colored lines indicate the separate
contributions proportional to
projections of $h_\parallel$, $h'_\parallel$, and $\nabla_\Omega^2 h_\parallel$.  
While for $l \lesssim 4$ all terms contribute significantly, 
the $h'_\parallel$ contribution dominates at higher $l$.  
This contribution is the same as the GW effect $\delta_h^s$ explored 
in \cite{bharadwaj/sarkar:2009}.  
We will return to this in \refsec{gal}.  Note
that the total contribution (black solid line) is significantly smaller
than the individual contributions for $l = 2,\,3$.  We will discuss this
in the next section.

%!!!!!!!!!!!!!!!!!!!!!!!!!!!!!!!!!!!!!!!!!!!
%!!!!!!!!!!!!!!!!!!!!!!!!!!!!!!!!!!!!!!!!!!!

%%%%%%%%%%%%%%%%%%%%%%%%%%%%%%%%%%%%%%%%%%%%%%%%%%%%%%%%%%%%%%%%%%%%%
\subsection{Quadrupole}
\label{sec:quad}

\reffig{Cl_TT} shows that the individual tensor contributions to
the observed angular power spectrum increase rapidly towards small $l$,
while the total contribution is much smaller.   These cancellations
are essentially a consequence of causality, which demands that the
observed clustering of galaxies cannot depend on tensor modes that
are super-horizon today.  

\begin{figure}[t!]
\centering
\includegraphics[width=0.5\textwidth]{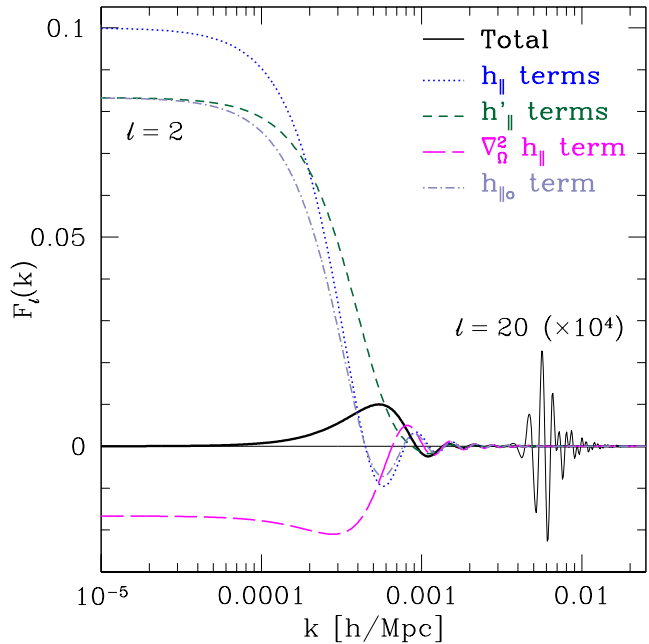}
\caption{Contributions to the kernel $F_l(k)$ for $l=2$ (thick) and
$l=20$ (thin, only total contribution shown, scaled by $10^4$), 
for a sharp source redshift
$\zt=2$ and the same parameters as in \reffig{Cl_TT}.  Note that the separate
contributions have non-zero weight for $k\to 0$, while the total $F^g_l$ 
(black solid) is only non-zero for $k\gtrsim 10^{-4}\iMpch$, as required
by causality.}
\label{fig:Fl}
\end{figure}

In particular, it is important to take into account the last term in 
\refeq{dgT2}, which quantifies the shearing of the observer's coordinate
system by the gravitational waves.  We refer to this as the observer term.  
Since $h_{ij}(o)$ is
a constant (transverse-traceless) tensor, this term only contributes to 
the quadrupole $l=2$.  
Specifically, we have an additional contribution to \refeq{Fl} for galaxies,
\ba
F_{l=2}^g(k) \to\:& F_{l=2}^g(k) + F_2^{g,o}(k) \vs
F_2^{g,o}(k) =\:& f_o \lim_{\chi\to 0} T_T(k,\eta_0-\chi) \frac{j_2(k\chi)}{(k\chi)^2} \vs
=\:& \frac1{15} f_o \,T_T(k,\eta_0).
\ea
The solid black line in \reffig{Cl_TT} includes this term, while the black
square indicates the value of $C_{l=2}$ we would obtain without this term.  
Neglecting the observer term results in an overestimation of the tensor
contribution to the galaxy quadrupole by three orders of magnitude.  The
reason for this significant effect becomes clear when considering 
the contributions to $F^g_{l=2}$ as function of $k$ (\reffig{Fl}).  
The individual contributions to $F^g_l$ approach a constant as $k\to 0$,
while the sum goes to zero for $k/H_0 \lesssim 1$, as demanded by
causality.  When neglecting the observer term (light blue in \reffig{Fl})
on the other hand, a residual constant contribution to $F^g_{l=2}$
remains for $k\to 0$, which together with the steeply falling tensor
power spectrum leads to a significant overestimation of the 
quadrupole.

%!!!!!!!!!!!!!!!!!!!!!!!!!!!!!!!!!!!!!!!!!!!
%!!!!!!!!!!!!!!!!!!!!!!!!!!!!!!!!!!!!!!!!!!!

%%%%%%%%%%%%%%%%%%%%%%%%%%%%%%%%%%%%%%%%%%%%%%%%%%%%%%%%%%%%%%%%%%%%%
\subsection{Limber approximation}
\label{sec:limber}

In the context of angular galaxy clustering, one often uses
the Limber approximation \cite{Limber} which significantly simplifies
the calculation of $C^g_l$.  The underlying assumption is that the
dominant contribution to the angular clustering comes from galaxy
pairs that are at similar distances along the line of sight.  It
is instructive to consider this approximation in the context of 
tensor modes.  Since the
Limber approximation works best for a broad redshift distribution,
we will here consider a redshift distribution roughly as expected for the 
Large Synoptic Survey Telescope (LSST \cite{LSST}),
\be
\frac{dN}{dz} \propto z^2 \exp\left[ -\left(\frac{z}{z_0}\right)^\beta\right],
\label{eq:dNdz}
\ee
with $z_0=0.15 $ and $\beta =0.73$, yielding a mean redshift of 1.2.  

The Limber approximation can formally be applied by using
\be
\frac2\pi \int k^2 dk\,  F(k,\chi)\,j_l(k\chi) j_l(k\chi') \approx
\frac{F\left(\frac{l+1/2}\chi,\chi\right)}{\chi^2} \d_D(\chi-\chi').
\ee
In the usual application to scalars, the functions of $k,\,\chi$ involved
(apart from the Bessel functions) are smooth and positive, whereas for 
tensor modes, the transfer function is oscillatory.  Performing the $k$ and 
then
one of the $\chi$ integrals in \refeq{CABl} leads to
\ba
C^{AB}_l \approx\:& \frac{(l+2)!}{(l-2)!} \frac{1}{(l+1/2)^4} \frac14
\int \frac{d\chi}{\chi^2} P_{T0}\left(\frac{l+1/2}{\chi}\right) \label{eq:CABLimber} \\
& \hspace*{0.7cm}\times \W_A(\chi) \W_B(\chi) 
\left[T_T\left((l+1/2) \frac{\eta_0-\chi}{\chi}\right)\right]^2.\nonumber
\ea
Here we have used that $T_T$ is only a function of $k\eta$.  
Note that for $l \gg 1$, the prefactor approaches $1/4$.  Given that 
$P_{T0} \propto k^{-3+n_T}$, we immediately see that the Limber approximation
predicts $C^g_l \propto l^{-3+n_T}$ for large $l$, if $\W_{A,B}$ are $l$-independent.  

\begin{figure}[t!]
\centering
\includegraphics[width=0.5\textwidth]{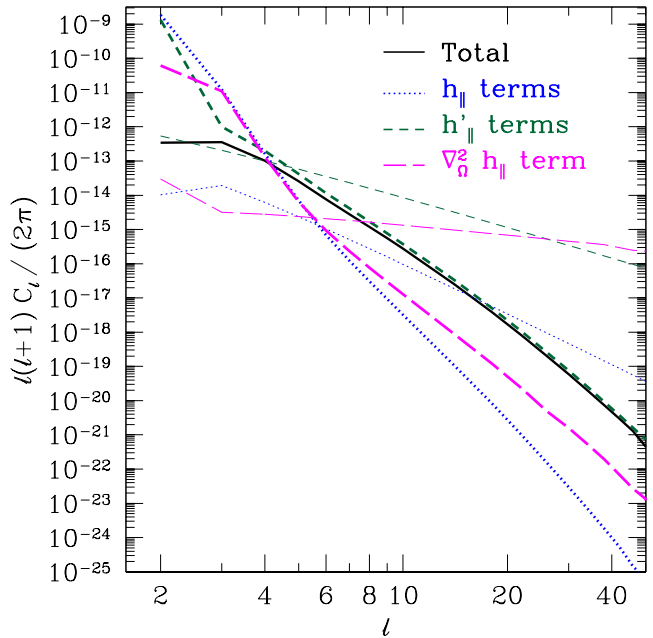}
\caption{Contributions of tensor modes to the angular galaxy power
spectrum for a broad redshift distribution (\refeq{dNdz}) expected
for LSST, separated into different contributions as in \reffig{Cl_TT}
(again using $b_e=2.5,\;Q=1.5$).  
The thick lines show the exact calculation, while the thin lines show
the Limber approximation using \refeq{CABLimber}.}
\label{fig:Limber}
\end{figure}

\reffig{Limber} shows the different contributions  to $C^g_l$ 
(as in \reffig{Cl_TT}) for the redshift distribution \refeq{dNdz} using
the full calculation (thick lines) and using the Limber approximation
(thin lines).  Clearly, the Limber approximation predicts the wrong
$l$-scaling of all terms, and we do not see any improvement in the 
approximation for high $l$ as in the scalar case.  Thus, the Limber
approximation is not applicable for calculating the tensor contribution
to the angular galaxy power spectrum at \emph{any} $\l$.  There are two 
reasons for this.  First, the tensor mode power spectrum is falling as 
$\sim k^{-3}$, so that the assumption that pairs of comparable line-of-sight
distance dominate because they have the smallest separation does not
hold.  Second, tensor modes oscillate and decay towards late times 
(while scalar modes
grow), so that the contributions to $C^g_l$ are concentrated at 
large scales and high redshifts near the source.

%!!!!!!!!!!!!!!!!!!!!!!!!!!!!!!!!!!!!!!!!!!!
%!!!!!!!!!!!!!!!!!!!!!!!!!!!!!!!!!!!!!!!!!!!

%%%%%%%%%%%%%%%%%%%%%%%%%%%%%%%%%%%%%%%%%%%%%%%%%%%%%%%%%%%%%%%%%%%%%
\subsection{Dependence on galaxy sample}
\label{sec:gal}

\begin{figure}[t!]
\centering
\includegraphics[width=0.5\textwidth]{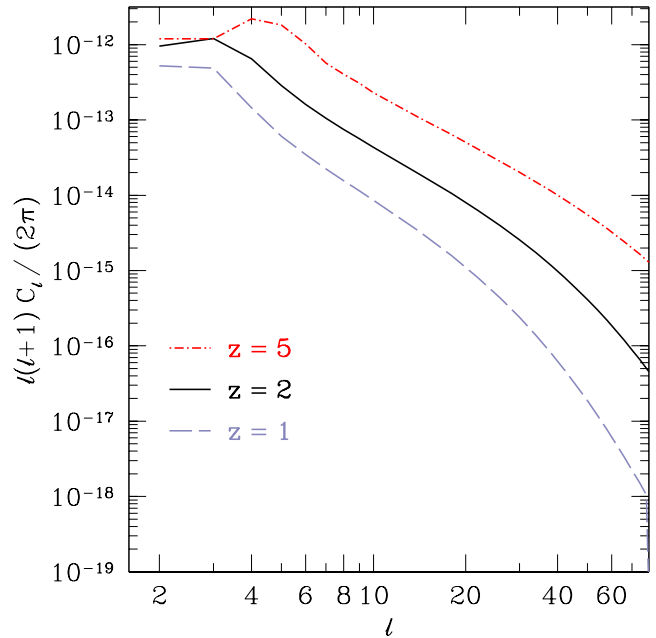}
\caption{Total tensor mode contribution to the angular galaxy power
spectrum as a function of source redshift, for a Gaussian redshift
distribution centered on $\zt$ with an RMS width of $0.03(1+\zt)$.  
Here, $b_e=2.5,\;Q=1.5$ fixed.}
\label{fig:Clz0}
\end{figure}

\reffig{Clz0} shows the tensor contribution to $C^g_l$ for different
source redshifts.  Here, we have assumed a Gaussian redshift distribution
centered on $\zt$ with RMS width of $0.03(1+\zt)$, emulating the effect
of photometric redshift errors.  We have kept $b_e=2.5,\;Q=1.5$ fixed.  
While the low multipoles ($l \lesssim 4$)
do not depend very strongly on $\zt$, the contribution for higher $l$ grows by
an order of magnitude when going from $\zt=1$ to $\zt=2$, and from
$\zt=2$ to $\zt=5$.  We found that reducing the width $\D\zt$ of the redshift 
distribution has no impact on the results at relevant $l$.  
The reason is that the bulk of the tensor mode contribution comes from 
near horizon-scale modes, as tensor modes decay once they enter the 
horizon.  Such long-wavelength modes are not affected by averaging
over a reasonably narrow redshift window ($\Delta\zt < 1$).  

\begin{figure}[t!]
\centering
\includegraphics[width=0.5\textwidth]{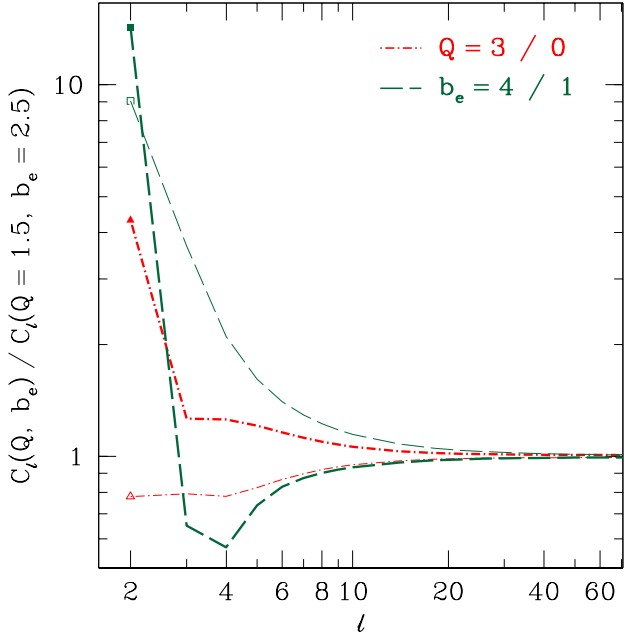}
\caption{Relative impact on the total tensor mode contribution to galaxy clustering
when changing $b_e$ from 2.5 to 4 and 1 (thick and thin green long-dashed,
respectively), and when changing $\Q$ from 1.5 to 3 and 0 (thick and thin
red dot-dashed, respectively).  Here, a Gaussian redshift distribution
centered on $\zt=2$ with RMS width $0.03(1+\zt)$ was assumed. 
}
\label{fig:ClbeQ}
\end{figure}

\reffig{ClbeQ} shows the effect of varying the galaxy sample parameters
$b_e$, quantifying the redshift evolution of the average number density,
and $\Q$, which determines the magnification bias contribution.  We see
a significant effect at low multipoles when varying $\Q$ and especially
$b_e$.  This is not surprising since we have seen that there is a high
degree of cancelation between different terms at low $l$ (Figs.~\ref{fig:Cl_TT} and
\ref{fig:Fl}), so that varying the coefficients of the different terms
can have a large impact.  

On the other hand, $C^g_l$ is insensitive to changes in $b_e$ and $\Q$
for $l\gtrsim 10$.  While surprising initially, this fact can be
understood straightforwardly.  At high $l$, the contributions from
$h'_\parallel$ dominate (\reffig{Cl_TT}).  Specifically, following
\refeq{dgT} there are two contributions, $f'_{\chit} h'_\parallel$ and
$f' \int h'_\parallel d\chi$.  The former contribution is a pure
line-of-sight volume distortion effect, and the coefficient 
$f'_{\chit} \propto H^{-1}(\zt)$ is independent of the galaxy sample.  
Now consider a single Fourier mode contributing to both terms.  
Neglecting factors of order unity,
we can estimate the ratio between the two terms as 
\ba
\frac{f' \int_0^{\chit} h'_\parallel d\chi}{f'_{\chit} h'_{\parallel}}
\sim\:& H(\zt) \frac{\int_0^{\chit} T'_T(k (\eta_0-\chi)) d\chi}{T'_T(k(\eta_0-\chit))},
\ea
where $T'_T = dT_T/d\eta$.  Since $T'_T$ decays rapidly once the argument
$k\eta$ become of order unity, we can approximate the numerator as
\be
\int_0^{\chit} T'_T(k (\eta_0-\chi)) d\chi \sim T'_T(k (\eta_0-\chit)) \Delta\chi,
\ee
where $\Delta\chi = \Delta(k\eta) / k \sim 1/k$ is the range in $\chi$
that contributes to the integral.  Thus, we arrive at an order-of-magnitude
estimate of
\ba
\frac{f' \int_0^{\chit} h'_\parallel d\chi}{f'_{\chit} h'_{\parallel}}
\sim\:& \frac{H(\zt)}{k}.
\label{eq:hpratio}
\ea
Physically, the integral of $h'_\parallel$ along the line of sight
leads to significant cancelations that become more severe as $k$ increases.  
It is not surprising then that the volume distortion term $f'_{\chit} h'_\parallel$
dominates at high $l$.  Specifically, using the kernel $F^g_l(k)$ for $l=20$
shown in \reffig{Fl}, we can estimate that the typical wavenumber of
tensor modes contributing to multipole $l$ is
\be
k_{\rm typ} \sim 0.003 \iMpch \times\frac{l}{10}.
\ee
\refeq{hpratio} then says that we expect $f'_{\chit} h'_\parallel$ to dominate
for $l \gtrsim 10$, in good agreeement with \reffig{ClbeQ}.  

Thus, the line-of-sight volume distortion $\propto h'_\parallel/H$ is
the single dominating term in $C^g_l$ at high $l$, which explains why the
tensor contribution is independent of the galaxy sample at high
multipoles.  Note however, that the prospects for detecting
this effect are mostly restricted to $l \lesssim 10$ anyway due to
the steep decline of the signal. 

%!!!!!!!!!!!!!!!!!!!!!!!!!!!!!!!!!!!!!!!!!!!
%!!!!!!!!!!!!!!!!!!!!!!!!!!!!!!!!!!!!!!!!!!!

%!!!!!!!!!!!!!!!!!!!!!!!!!!!!!!!!!!!!!!!!!!!
%!!!!!!!!!!!!!!!!!!!!!!!!!!!!!!!!!!!!!!!!!!!

%!!!!!!!!!!!!!!!!!!!!!!!!!!!!!!!!!!!!!!!!!!!
\begin{figure}[t!]
\centering
\includegraphics[width=0.5\textwidth]{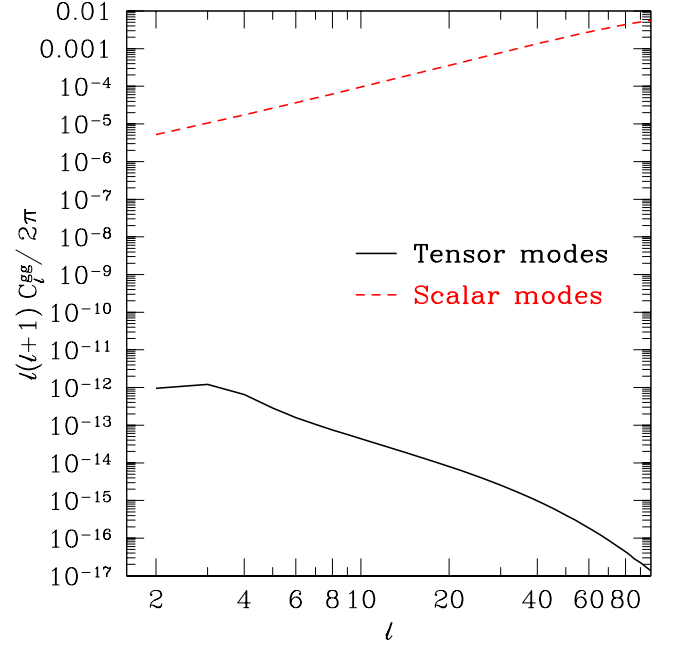}
\caption{Comparison between tensor mode contributions to the 
galaxy power spectrum (black solid) for $\zt=2$ (cf. \reffig{Clz0}),
and scalar contributions for a linear bias $b=2$ (see \refapp{Clscalar}
for details on the calculation of the scalar contributions).  
}
\label{fig:Clcomp}
\end{figure}
%!!!!!!!!!!!!!!!!!!!!!!!!!!!!!!!!!!!!!!!!!!!

%%%%%%%%%%%%%%%%%%%%%%%%%%%%%%%%%%%%%%%%%%%%%%%%%%%%%%%%%%%%%%%%%%%%%
%\subsection{Cross-correlation of different redshift slices}
%\label{sec:cross}

%%%%%%%%%%%%%%%%%%%%%%%%%%%%%%%%%%%%%%%%%%%%%%%%%%%%%%%%%%%%%%%%%%%%%
%%%%%%%%%%%%%%%%%%%%%%%%%%%%%%%%%%%%%%%%%%%%%%%%%%%%%%%%%%%%%%%%%%%%%
\section{Higher order terms}
\label{sec:2nd}

So far, we have only kept terms linear in all perturbations.  
While all terms second and higher order in $h_{ij}$ are likely
irrelevant given the small amplitude of tensor perturbations,
there are terms of order $h_{ij}\,\d_g$ which can be much less suppressed
on small scales where $\d_g$ can become order unity.  There are
more terms of order $h_{ij}\, A$ where $A$ stands for any scalar
metric perturbation.  However, those terms are much smaller than
$h_{ij}\,\d_g$ and can be neglected.  Note that as long as scalar
and tensor modes do not correlate, the lowest non-trivial statistic
induced by these terms is a four-point function of $\tilde\d_g$.  
A specific quadratic estimator for the tensor modes can be constructed
based on these terms \cite{MasuiPen,BookMKFS}.  Physically, one uses
the anisotropy of the small-scale correlation function of $\tilde\d_g$
to search for coherent large-scale distortions induced by tensor
modes.  Most of the signal-to-noise for detecting a GW background
is contained on very small scales \cite{BookMKFS}.  

Noting that $\vx = \tilde\vx-\D\vx$ and $1+\bar z = (1+\zt)(1-\d z)$,
the relevant terms neglected in going from \refeq{cov_dens} to \refeq{dT1}
are given by
\ba
\tilde\d_{gT,\rm 2nd} =\:& \d_g\left[\tilde\vx-\D\vx,(1+\zt)(1-\d z)-1\right]
- \d_g\left(\tilde\vx,\zt\right) \vs
=\:& -\D\vx\cdot\v{\nabla}\d_g(\tilde\vx,\zt) - \d z \frac{\partial}{\partial\ln (1+\zt)} \d_g(\tilde\vx,\zt),
\label{eq:2nd}
\ea
where $\d_g$ is the intrinsic fractional perturbation in the comoving
galaxy density.  
Note that both $\D\vx$ and $\d z$ are linear in $h_{ij}$.  If we 
approximate $\nabla\d_g \sim \d_g/r$, where $r$ is the scale on which 
the correlation function of $\d_g$ is measured, then the second
term is suppressed with respect to the first term by a factor of $r/\chit$,
since $\partial\ln\d_g/\partial\ln(1+z)$ is typically order unity.  
This term is thus highly suppressed on small scales where the second-order 
terms become relevant.  While the intrinsic two-point function $\xi(r)$ of the
tracer is isotropic and location-independent (neglecting redshift-space,
tidal distortions and intrinsic anisotropy/inhomogeneity
\cite{MasuiPen,jeong/kamionkowski:2012}), 
the observed correlation function
\be
\tilde\xi(\vnhat,\zt;\vnhat',\zt') = \< \tilde \d(\tilde\vx) \tilde\d(\tilde \vx')\>
\ee
is anisotropic and depends on the location $\tilde\vx$.  This can be used
to measure the distortions by tensor perturbations.  Specifically, since
we have 6 free parameters in $\tilde\vx,\,\tilde\vx'$, we can measure
6 components of the distortion field.  At this point, it is important to
stress that the terms in \refeq{2nd} are \emph{not observable} directly
(as a simple example, consider the case of a constant deflection 
$\D\vx=$~const).  In order to determine which quantities are actually 
observable, consider contours of constant $\tilde\xi$,
\be
\tilde\xi(\vnhat,\zt;\vnhat,\zt') = \xi_0.
\ee
These contours correpond to a fixed physical scale $r_0$ (on a 
constant-proper-time hypersurface) at the source through
\be
\xi(r_0) = \xi_0.  
\ee
In other words, the intrinsic homogeneous and isotropic correlation function
$\xi(r)$ is supplying us with a ``standard ruler'' $r_0$ (or, a set of 
standard rulers as we are free to vary $\xi_0$).  In \citet{stdruler},
we carefully define a general standard ruler and derive the properties
of the deflection field which are observable through it.  As we have
seen here, the distortion of correlation functions by tensor modes is
one application of the results of \cite{stdruler}.  

Finally, we point out that a non-zero three-leg coupling 
$\left<\delta(\bm{k}_1)\delta(\bm{k}_2) h_{ij}(\bm{k}_3)\right>$ present 
at an early 
stage of the universe can also imprint its signature as a local departure from
statistical homogeneity.  The optimal estimator for the amplitude of tensor 
perturbations given such a coupling has been 
constructed in \cite{jeong/kamionkowski:2012}.

%!!!!!!!!!!!!!!!!!!!!!!!!!!!!!!!!!!!!!!!!!!!
\begin{figure}[t!]
\centering
\includegraphics[width=0.5\textwidth]{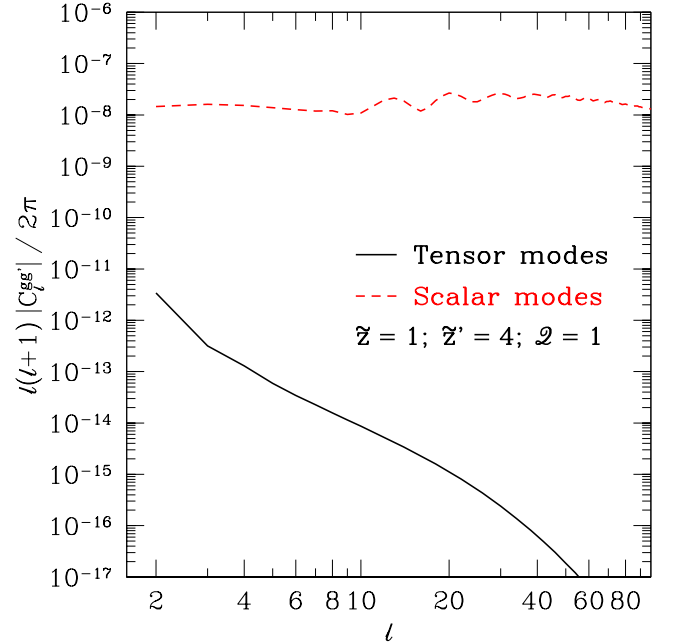}
\caption{Comparison between tensor mode and scalar contributions to the 
angular cross-correlation between two widely separated redshift bins
($\zt = 1,\, \zt' = 4$) and for $\Q=1$ so that most magnification
contributions drop out (linear bias $=2$; other parameters as 
in \reffig{Clz0}).  
}
\label{fig:Clcrosscomp}
\end{figure}
%!!!!!!!!!!!!!!!!!!!!!!!!!!!!!!!!!!!!!!!!!!!
%%%%%%%%%%%%%%%%%%%%%%%%%%%%%%%%%%%%%%%%%%%%%%%%%%%%%%%%%%%%%%%%%%%%%
%%%%%%%%%%%%%%%%%%%%%%%%%%%%%%%%%%%%%%%%%%%%%%%%%%%%%%%%%%%%%%%%%%%%%
\section{Discussion}
\label{sec:concl}

We have derived the complete tensor contributions to the observed
galaxy density at linear order.  The result is summarized in 
\refeq{dgT}.  At this order, gravitational waves do not perturb the
intrinsic physical density of tracers; thus all contributions are
projection effects from the effects of GW on the propagation of light.  
We have found that, contrary to gravitational lensing by scalar perturbations,
tensor perturbations contribute mainly at redshifts close to the source
redshift.  Together with the scale-invariant power spectrum of GW, this 
results in a steeply falling angular power spectrum of the
tensor contributions, with multipoles $l \sim 10$ already being suppressed
by an order of magnitude with respect to $l=2 - 4$.  

\reffig{Clcomp} shows a comparison of the tensor contributions with the
scalar contributions to the galaxy density.  Here, we have assumed a
linear bias of $b = 2$, and all relativistic corrections are included
following \cite{gaugePk} (see \refapp{Clscalar} for details).  
Clearly, the tensor
contributions are suppressed by $\sim$7 orders of magnitude
with respect to the scalar contributions at the largest scales, for the
maximum currently allowed value of $r=0.2$.  
One might wonder whether galaxy \emph{cross}-correlations, i.e. between
different redshift bins, could be more promising.  After all, when
cross-correlating widely separated redshift slices, there is little
intrinsic correlation of galaxies, and the main contribution comes
from lensing (magnification bias effect $\propto 2(\Q-1)\hat\k$).  Hence, the 
most optimistic case for detecting tensor modes through their modulation
of the galaxy density would consist of cross-correlating two galaxy 
populations widely separated in redshift, both of which have $\Q=1$ so
that the magnification bias effect drops out (this could be achieved,
for example, by selecting galaxies on surface brightness).  
This most optmistic case is shown in \reffig{Clcrosscomp}.  We find that 
the tensor contribution can become
as large as $10^{-3}$ times the residual scalar contribution; however,
this is still much too small to be detectable.  

Thus, given that we are not able to directly distinguish between scalar 
and tensor contributions to the galaxy angular power spectrum, we
do not expect much detection potential for gravitational waves from
the leading order effect on the angular power spectrum of galaxies.  
However, there are terms of order $h_{ij}\,\d_g$ (where $\d_g$ is the
intrinsic galaxy overdensity) which induce a particular four-point 
correlation function in the observed galaxy density $\tilde\d_g$.  
This is the effect exploited in \cite{MasuiPen,BookMKFS} for projected
constraints from 21cm emission from the dark ages.  In essence, 
the intrinsic (homogeneous and isotropic) correlation function
provides us with a standard ruler that allows us to observe certain
properties of the distortion field, as derived in \cite{stdruler}.  
Thus, one can apply the
the scalar-vector-tensor decomposition on the sky described in \cite{stdruler}.  
Most importantly, both the (2-)vector and (2-)tensor components allow for 
an $E/B$-decomposition, so that any scalar contributions to the distortion
(e.g. from lensing or redshift-space distortions) do not contribute 
to the $B$-mode at linear order.  We leave a more detailed
investigation of these possibilities for future work.

%%%%%%%%%%%%%%%%%%%%%%%%%%%%%%%%%%%%%%%%%%%%%%%%%%%%%%%%%%%%%%%%%%%%%%%%%%%
\acknowledgments
We would like to thank Chris Hirata and Marc Kamionkowski for helpful
discussions.  FS is supported by the Gordon and Betty Moore Foundation 
at Caltech. DJ was supported by NASA NNX12AE86G.

%%%%%%%%%%%%%%%%%%%%%%%%%%%%%%%%%%%%%%%%%%%%%%%%%%%%%%%%%%%%%%%%%%%%%%%%%%%
%%%%%%%%%%%%%%%%%%%%%%%%%%%%%%%%%%%%%%%%%%%%%%%%%%%%%%%%%%%%%%%%%%%%%%%%%%%
%\clearpage
\appendix

%%%%%%%%%%%%%%%%%%%%%%%%%%%%%%%%%%%%%%%%%%%%%%%%%%%%%%%%%%%%%%%%%%%%%%%%%%%
\section{Convergence}
\label{app:kappa}

We define the convergence through the  
divergence of the transverse displacements:
\ba
\hat\k \equiv\:& -\frac12 \partial_{\perp\,i}\Delta x_\perp^i 
\ea
First, taking the transverse divergence of \refeq{Dxperp} yields
\ba
-2 \hat\kappa =\:& \frac12 \chit \left( h^i_{\;j} \partial_{\perp i} \nhat^j 
- h_\parallel \partial_{\perp i} \nhat^i\right)_o \\
&  + \int_0^{\chit} d\chi \Big[\! - \partial_{\perp i} \left( h^i_{\;j}\nhat^j 
- h_\parallel \nhat^i\right) \vs
& \hspace*{1.5cm}
+ \frac12 (\chit-\chi) \frac{\chi}{\chit} \nabla_\perp^2 h_\parallel
\Big].\nonumber
\ea
Using the fact that $h_{ij}$ is transverse and that 
$\partial_\chi = \partial_\parallel - \partial_\eta$,
straightforward algebra then yields
\ba
\hat\kappa =\:& \frac54 h_{\parallel o} -\frac12 h_{\parallel}
 - \frac12 \int_0^{\chit} d\chi \Big[ 
h_\parallel' + \frac3{\chi} h_\parallel \Big] \vs
& - \frac14 \nabla^2_\Omega \int_0^{\chit} d\chi \frac{\chit-\chi}{\chit\:\chi} 
h_\parallel. \label{eq:kappa2}
\ea
The last term is familiar as the one dominating on small scales
(Newtonian limit) for the scalar case, and is shown separately
as blue dotted lines in \reffig{Cl_TT} and \reffig{Limber}.   Note that 
due to the very different power spectrum of tensor modes, and the
fact that tensor modes decay with decreasing redshift, this term 
actually becomes suppressed on small scales.

%%%%%%%%%%%%%%%%%%%%%%%%%%%%%%%%%%%%%%%%%%%%%%%%%%%%%%%%%%%%%%%%%%%%%%%%%%%
\section{Angular galaxy power spectrum from scalar
perturbations}
\label{app:Clscalar}

In this section, we provide expressions for the angular power spectrum
of galaxies due to scalar perturbations (see \reffig{Clcomp} and \ref{fig:Clcrosscomp}), including all 
relativistic corrections
as derived in \cite{yoo/etal:2009,challinor/lewis:2011,bonvin/durrer:2011,gaugePk}.  

It is convenient to use expressions in conformal-Newtonian gauge,
where we write the metric as
\be
ds^2 = a^2(\eta)\left[- (1+2\Psi) d\eta^2
+ (1+2\Phi) \d_{ij} dx^i dx^j \right].
\label{eq:metric_cN}
\ee
Further, for convenience we write the velocity $v_i$ in terms of a
scalar velocity potential $V$, $v_i = V_{,i}$, and relate $V,\,\Phi,\,\Psi$
to the density contrast $\d_m^{\rm sc}$ in synchronous-comoving gauge
through (see \cite{Schmidt08,HuSaw07b})
\ba
V(\vk,\eta) =\:& a H f k^{-2} D(a(\eta)) \d_m^{\rm sc}(\vk,\eta_0)
\vs
\Phi(\vk,\eta) -\Psi(\vk,\eta) =\:& D_{\Phi_-}(k,\eta) \d_m^{\rm sc}(\vk,\eta_0) \vs
\Phi(\vk,\eta) + \Psi(\vk,\eta) =\:& g(k,\eta) D_{\Phi_-}(k,\eta) \d_m^{\rm sc}(\vk,\eta_0) \vs
\Rightarrow \Phi(\vk,\eta) =\:& \frac12 [g+1] D_{\Phi_-} \d_m^{\rm sc}(\vk,\eta_0) \vs
 \Psi(\vk,\eta) =\:& \frac12 [g-1] D_{\Phi_-} \d_m^{\rm sc}(\vk,\eta_0) \vs
(\Phi-\Psi)'(\vk,\eta) =\:& D_{\rm ISW}(k,\eta) \d_m^{\rm sc}(\vk,\eta_0),
\ea
where $f \equiv d\ln D/d\ln a$, $D(a)$ is the matter growth factor
(normalized to unity at $a=1$) and we have defined general coefficient
functions $D_{\Phi_-},\,g,\,D_{\rm ISW}$ to allow for non-standard 
cosmologies.  In a $\Lambda$CDM cosmology, we have
\ba
D_{\Phi_-}(\vk,\eta) =\:& 
3\Omega_m \frac{a^2H^2}{k^2}D(a(\eta))
=
3 \Omega_{m0} \frac{H_0^2}{k^2} a^{-1}(\eta) D(a(\eta))\vs
g(\vk,\eta) =\:& 0.
\ea
Here, the subscript $0$ denotes that the quantity is evaluated at $z=0$.
Throughout, we will drop observer terms
corresponding to the monopole and dipole, but keep terms that
contribute to the quadrupole.  

Converting Eqs.~(55) and (72) in \cite{gaugePk} from synchronous-comoving
gauge to conformal-Newtonian gauge, using the relations given in App.~D
of \cite{gaugePk}, we obtain (see also \cite{challinor/lewis:2011})
\ba
\tilde{\delta}_g(\tilde\vx)
=\:& \d_g + b_e(\d z - a H V)
- \frac1{a H} \partial_\parallel^2 V
\vs
&- \left(
1 - \frac{1}{aH}\frac{d H}{d z}
+
\frac{2}{a H \chit}
\right)_{\zt} [\d z - a H V] \vs
& + 2(\Phi + a H V )
\vs
& -\frac{2}{\chit} V
-\frac{2}{\chit}
\int_0^{\chit}d\chi
(\Phi-\Psi) - 2\k
\vs &
-\frac{1}{aH}\left(\Phi + a H V\right)'
+ \Q \M \vs
%%%
\M =\:&
-2\Phi - 2a H V + \frac2{\chit} V + 2\k 
\vs
& + \frac2{\chit} \int_0^{\chit}(\Phi-\Psi)d\chi 
+ 2\left[\frac1{a H \chit} - 1\right]
(\d z - a H V),\nonumber
\ea
where
\ba
\d z =\:& \partial_\parallel V - \Psi + \int_0^{\chit} d\chi (\Phi-\Psi)' \vs
\k =\:& -\frac12 \int_0^{\chit} d\chi (\chit-\chi)\frac{\chi}{\chit} \nabla_\perp^2
(\Phi-\Psi)
\ea
denote the redshift perturbation and convergence, respectively, 
in conformal-Newtonian gauge.  Assuming a linear bias relation in
synchronous-comoving gauge (as discussed in \cite{gaugePk}), and expanding 
the different contributions, we obtain
\ba
\tilde{\delta}_g(\tilde\vx) =\:& b \d_m^{\rm sc} \vs
& + \left[b_e - 1 - 2\Q + \frac1{aH}\frac{dH}{dz} + 2(\Q-1) \frac1{aH\chit}\right]\d z \vs
& + \left[ (3-b_e) aH - \frac{dH}{dz}\right] V + 2(1-\Q) \Phi \vs
& + 2(\Q-1) \int_0^{\chit}\frac{d\chi}{\chit} (\Phi-\Psi) 
+ 2(\Q-1) \k \vs
& - \frac1{aH}\left[\partial_\parallel^2 V + (\Phi+ aH V)'\right].
\ea
We can now evaluate the contribution of a single plane wave 
along the $z$-axis, i.e. $\d^{\rm sc}_m(\vx,\eta) = \d^{\rm sc}_m(\vk,\eta_0) D(a(\eta)) e^{i \vx\vk}$.  
Further, we write $\vx = \vnhat \chi$, and $\vx\cdot\vk = x \mu$ 
with $x=k\chi$,
$\tilde x = k\chit$.  
First, we have
\begin{widetext}
\ba
\d z =\:& \d_m^{\rm sc}(\vk, \eta_0)
\left[ \left( \frac{a H f D}{k}\partial_{\tilde x} - \frac12(g-1) D_{\Phi_-} \right)_{\zt} e^{i\tilde x\mu}
 + \int_0^{\chit} d\chi D_{\rm ISW} e^{i x \mu}\right]
\vs
\k =\:& \frac12 l(l+1) \int_0^{\chit} d\chi \frac{\chit-\chi}{\chi\chit} D_{\Phi_-}(k,\eta_0-\chi) e^{ix\mu} \: \d_m^{\rm sc}(\vk,\eta_0) \vs
\partial_\parallel^2 V =\:& ( a H f D)_{\zt}\,\partial_{\tilde x}^2 e^{i\tilde x\mu}\, \d_m^{\rm sc}(\vk,\eta_0)
\vs
\Phi' =\:& \frac12\left( g' D_{\Phi_-} + [g+1] D_{\rm ISW}\right)_{k,\tilde\eta}
e^{i\tilde x\mu} \d_m^{\rm sc}(\vk,\eta_0)
\vs
\frac1{aH} (a H V)' =\:& \left(\frac{aH}{k}\right)_{\zt}^2 (f D)_{\zt} \left[2\frac{d\ln a H}{d\ln a} + \frac{d\ln f D}{d\ln a}\right]_{\zt} e^{i\tilde x \mu} \d_m^{\rm sc}(\vk,\eta_0).
\ea
Following the procedure described in App.~A1 of \cite{stdruler}, it
is then straightforward to derive the angular power spectrum
of $\tilde \d_g$, for a sharp source redshift $\zt$, in terms of the
matter power spectrum today $P_m(k)$ in synchronous-comoving gauge:
\ba
C_{\tilde g}(l) =\:& \frac2\pi \int k^2 dk\,P_m(k) |F_l^{\tilde g}(k)|^2 
\vs
F_l^{\tilde g}(k) =\:& b D(\tilde a) j_l(\tilde x) 
+ \left[b_e - 1 - 2\Q + \frac1{aH}\frac{dH}{dz} + 2(\Q-1) \frac1{aH\chit}\right]_{\zt}\,
F_l^{\d z}(k) \vs
&  + \left[ (3-b_e) - \frac1{aH}\frac{dH}{dz}\right]_{\zt}\, \left(\frac{aH}{k}\right)_{\zt}^2 (f D)_{\zt}\,
j_l(\tilde x)
+ 2(1-\Q) \frac12 ([g+1] D_{\Phi_-})_{\zt}\, j_l(\tilde x) \vs
& + 2(\Q-1) \left[\int_0^{\chit}\frac{d\chi}{\chit} D_{\Phi_-} j_l(x) 
+
\frac12 l(l+1) \int_0^{\chit} d\chi \frac{\chit-\chi}{\chi\chit} D_{\Phi_-} j_l(x) \right] - (f D)_{\zt}\, \partial_{\tilde x}^2 j_l(\tilde x) 
\vs
&  - \frac1{aH} \frac12\left( g' D_{\Phi_-} + [g+1] D_{\rm ISW}\right)_{\zt}\, j_l(\tilde x) 
 - \left(\frac{aH}{k}\right)_{\zt}^2 (f D)_{\zt}\, \left[2\frac{d\ln a H}{d\ln a} + \frac{d\ln f D}{d\ln a}\right]_{\zt}\, j_l(\tilde x)
\label{eq:Cldg}\\
%%%
F_l^{\d z}(k) =\:&
\left(\frac{a H f D}{k}\partial_{\tilde x} - \frac12(g-1) D_{\Phi_-} \right)_{\zt} j_l(\tilde x)
 + \int_0^{\chit} d\chi D_{\rm ISW} j_l(x),
\label{eq:Fldz}
\ea
where again $x = k\chi,\, \tilde x = k\chit$.  We can also
easily derive the power spectrum of the magnification itself, which is
an observable (see \cite{stdruler}):
\ba
C_{\M}(l) =\:& \frac2\pi \int k^2 dk\,P_m(k) |F_l^{\M}(k)|^2 
\vs
F_l^{\M}(k) =\:& 2\left[\frac1{a H \chit} - 1\right] F_l^{\d z}(k) 
- ([g+1] D_{\Phi_-})_{\zt}\, j_l(\tilde x) \vs
& + l(l+1) \int_0^{\chit} d\chi \frac{\chit-\chi}{\chi\chit} D_{\Phi_-} j_l(x) 
+ 2 \int_0^{\chit}\frac{d\chi}{\chit} D_{\Phi_-} j_l(x).
\label{eq:Clmag}
\ea
These expressions are for a sharp source redshift.  It is straightforward
to generalize them to a distribution $dN/d\zt$ of source redshifts
(where $dN/d\zt$ is assumed normalized), following the discussion in
\refsec{ClA}.    
In particular, contributions to $F^X_l(k)$ of the form
\be
\int_0^{\chit} d\chi\,W(\chit, \chi) \hat Q(x) j_l(x),
\ee
where $\hat Q(x)$ is any derivative operator, are generalized to
\ba
\int_0^{\chit} d\chi\,W(\chit, \chi) \hat Q(x) j_l(x) \to\:&
\int_0^{\infty} d\chi\,\W(\chi) \hat Q(x) j_l(x) \vs
\W(\chi) =\:& \int_{z(\chi)}^\infty d\zt\,\frac{dN}{d\zt} W(\chit=\chib(\zt), \chi).
\ea
Further, contributions evaluated at the source are generalized to
\ba
A(\chit) \hat Q(\tilde x) j_l(\tilde x) \to\:&
\int_0^{\infty} d\chi\,\W(\chi) \hat Q(x) j_l(x) \vs
\W(\chi) =\:& \left[\frac{dN}{dz} H(z)\right]_{z(\chi)}\!\! A(\chi).
\ea
\end{widetext}

%%%%%%%%%%%%%%%%%%%%%%%%%%%%%%%%%%%%%%%%%%%%%%%%%%%%%%%%%%%%%%%%%%%%%%%%%%%
%\bibliographystyle{arxiv_physrev}
\bibliography{GW}

\end{document}